# Integrating Over Sea Radio Channel for Sea Turtles Localization in the Indian Ocean

Loic Guegan[1], Nour Mohammad Murad[1], *Member, IEEE,* Jean Mickael Lebreton[1] and Sylvain Bonhommeau[2]
Laboratory of Energetic, Electronic and Processes (LE2P)
1: University of La Reunion, 2: IFREMER
Saint-Pierre, France
Email: loicguegan@loicguegan.fr, nour.murad@univ-reunion.fr

*Abstract*—This paper deals with the modeling of over sea radio channel with the aim of establishing sea turtle localization off the coast of Reunion Island but also on Europa Island in the Mozambique Channel. In order to model this radio channel, we are making a measurement protocol. In a first approach, measurements of turtle trajectory were done over land and finally it will be conducted over sea. We have scheduled an over sea measurement campaign in the middle of June. This paper shows a signal cross correlation technique used to characterize the over sea propagation channel.

## I. Introduction

We were solicited by French Research Institute for Exploitation of the Sea [1] to set up a localization system of sea turtles. Indeed, there is a real need to know the position of marine turtles in order to be able to effectively study their behaviors and determine their foraging habitat. There are different GPS modules for the localization of sea turtles available on the market, but they are most often expensive and researchers have no access to the algorithms used to estimate animal positions. To overcome these disadvantages, we have chosen to use several radio modules, associated with different localization algorithms [2] that involve gateways to be placed on the coast. The distances between sea turtles and gateways (GW) will be in the order of kilometers, which is acceptable for this type of localization. Juvenile turtles that we focus the study on are generally foraging in coastal areas.

Sea turtles can reach peak speeds of the order of 35 km/h and can have very short times spent at the surface to breathe (between 100 and 500 milliseconds) and more longer times (2 to 3 minutes) [3]. In addition, it is necessary to take into account the radio conditions of the over sea environment, which may be extreme in some cases [4,5]. Wind, waves and sea spray are factors that can alter the signal. This is why we have decided to model the over sea radio channel conditions, in order to adapt the RF technologies and the algorithms used to attain the best possible accuracy on turtle localization.

## II. Measurement approach

To model the over sea channel, we initially made a terrestrial measurement protocol composed of a transmitter representing the turtle and a receiver representing a GW. We are using Software Define Radio to be able to perform an easy and fast post-processing on the computer. Also, we configured a transmitter and a receiver at a carrier frequency $fc = 868 MHz$ like the LoRa European Band and a transmit power $Tx = 9 dBm$ for realizing a power profile of the radio channel. We defined a specific sequence call $Seq$ with a length $N = 456 bits$ and a sample rate $f_r = 10 MHz$. Thanks to this, we are able to have a path resolution channel in the order of $100 ns$. On the receiver side, a cross correlation algorithm is implemented along an observation window $W$ to extract some channel profiles using a specific modulation.

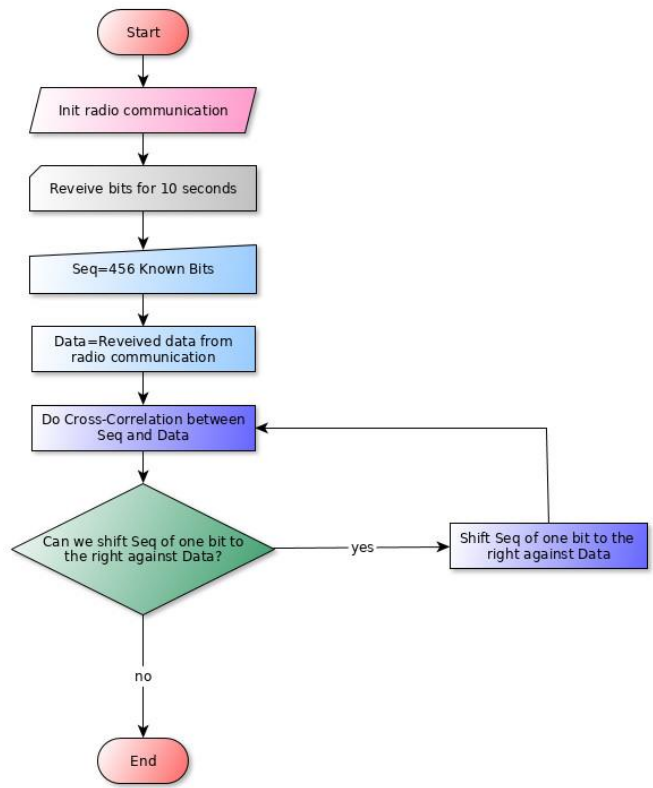

Fig. 1. Flowchart of the algorithm used on the receiver in order to realize a channel profile.

## III. Analysis of the First results

Different experimental scenarios are done with the strong hypothesis that the Turtle Transmitter (TT) is static for the moment and always on the ground. Different initial conditions are varied in order to analyze their influence on the RF signal propagation. First of all, the height $h$ of the GW is varied in order to study its influence on the reception quality (multi-paths, ...). Then, we changed the distance $d$ between the TT and the GW in order to highlight the signal attenuation.

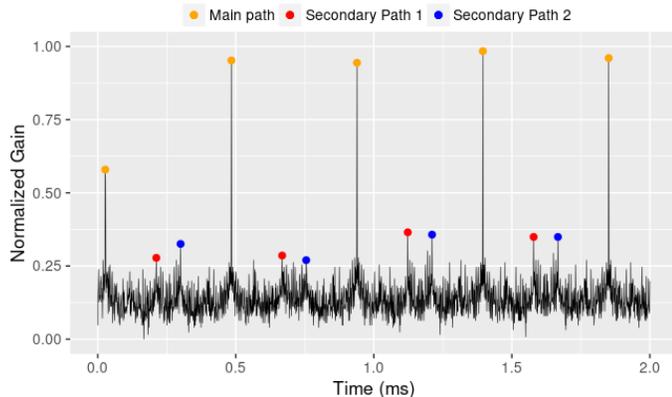

Fig. 2. Normalized channel profile at d=20m, h=1m and W=2ms (scenario 1).

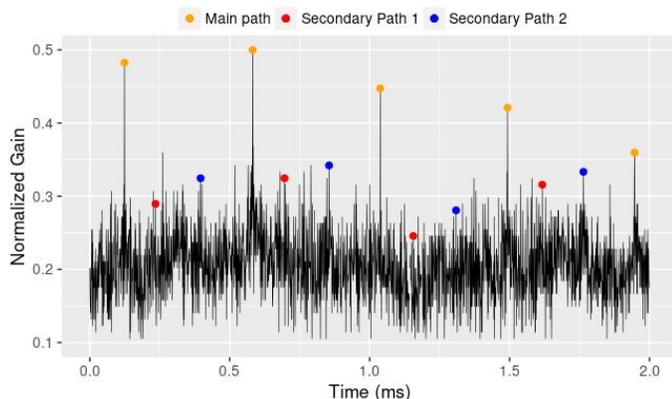

Fig. 3. Normalized channel profile at d=20m, h=0m and W=2ms (scenario 2).

The results obtained by these two scenarios highlight the importance of the GW height. Indeed, the reflection of the different signal paths on the ground attenuates the main signal path in a significant way. A more precise analysis of the results of scenario 1 and 2 shows that different secondary paths occurred and disturbed the main signal path. The table I shows the channel profiles of the different scenarios.

TABLE I. CHANNEL PROFILE OF SCENARIO 1 AND 2.

|  | Scenario 1 | | Scenario 2 | |
| --- | --- | --- | --- | --- |
| Paths | Delay (µs) | Mean Gain (dB) | Delay (µs) | Mean Gain (dB) |
| 1 | 0 | -0,38 | 0 | -3,69 |
| 2 | 18,38 | -5,00 | 11,73 | -5,36 |
| 3 | 27,19 | -4,74 | 26,89 | -5,14 |

The sequence $Seq$ is not encoded to protect the data and it is transmitted directly in order to consume the least amount of energy as possible to preserve battery life on the TT. And for the same reason we cannot increase the transmit power $Tx$.

## IV. SEA TURTLE LOCALIZATION

The figure 4 presents the localization results realized on the Omnet++ simulator and combined with Matlab to simulate the propagation channel. On the left side of the figure 4 the localization is made by using an ideal channel called *FreePathLoss* and on the right side with the realistic channel model of the scenario 2. The algorithm used here is the Power Of Arrival (POA). Other algorithms are already implemented like the Time Of Arrival (TOA) and the Time Difference Of Arrival (TDOA). The implementation of the scenario 2 channel gives more realistic results and this is why we are planning to integrate the over sea channel model.

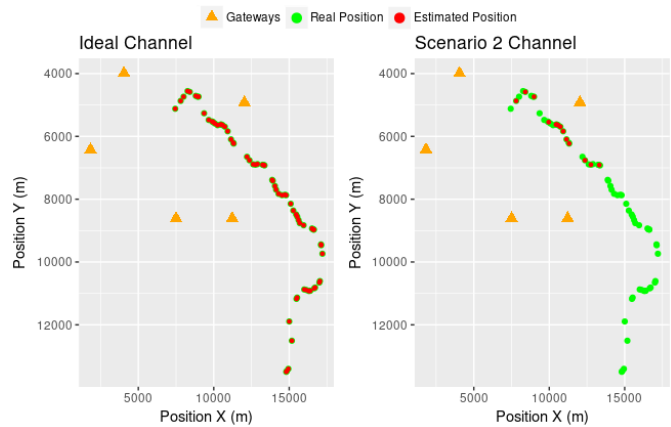

Fig. 4. Localization of a turtle by trilateration using POA on ideal channel model and scenario 2 channel model.

## V. CONCLUSION

This paper shows the approach used to model an RF channel in real conditions with a resolution of $100ns$. This approach has been validated for terrestrial purpose based on two parameters, the height and the distance. The height of the different gateways on the coast will be very important for localization accuracy. After this integration of the over land radio channel, we are planning to carry out measurement campaigns in order to be able to characterize the over sea radio channel. In this way, we will take into account the effects of the radio channel on the localization accuracy.